\documentclass[twocolumn,prl]{revtex4}
\usepackage{stmaryrd}
\usepackage{marvosym}
\usepackage{url}
\usepackage{subfigure}
\usepackage{amsfonts}
\usepackage{amsmath}
\usepackage{amssymb}
\usepackage{dcolumn}
\usepackage{bm}
\usepackage{epsfig}
\usepackage{graphicx,graphics}
\usepackage{wrapfig}
\usepackage{dcolumn}
\usepackage{fancybox}
\usepackage{euscript}
\usepackage[german,english]{babel}
\usepackage{psfrag}
\usepackage{color}
\usepackage[usenames,dvipsnames]{xcolor}

\newcommand{\ket}[1]{\left|{#1}\right\rangle}
\newcommand{\bra}[1]{\left\langle{#1}\right|}

\begin{document}

\title{Optical switching of radical pair conformation enhances magnetic sensitivity}
\author{Gian Giacomo Guerreschi$^{1,2}$, Markus Tiersch$^{1,2}$, Ulrich E. Steiner$^3$, and Hans J. Briegel$^{1,2}$}
\affiliation{$^1$Institut f\"ur Quantenoptik und Quanteninformation der
\"Osterreichischen Akademie der Wissenschaften, Innsbruck, Austria\\
$^2$Institut f{\"u}r Theoretische Physik, Universit{\"a}t Innsbruck,
Technikerstra{\ss }e 25, A-6020 Innsbruck, Austria\\
$^3$Fachbereich Chemie, Universit\"at Konstanz, D-78457 Konstanz, Germany}

\begin{abstract}
The yield of chemical reactions involving intermediate radical
pairs is influenced by magnetic fields well beyond the levels expected
from energy considerations.
This dependence can be traced back to the microscopic dynamics of
electron spins and constitutes the basis of the chemical compass.
Here we propose a new experimental approach based on molecular
photoswitches to achieve additional control on the chemical reaction
and to allow short-time resolution of the spin dynamics.
Our proposal enables experiments to test some of the standard assumptions
of the radical pair model and improves the sensitivity of chemical
magnetometers by two orders of magnitude.
\end{abstract}

\maketitle

\textit{Introduction.---}
Radical pair (RP) reactions have recently raised attention as one of the best
characterized instances in the field of quantum biology.
Due to the strong dependence of the products' chemical yield on the external
magnetic field, they constitute one of the two main hypotheses to explain how certain
species of birds use the Earth's magnetic field to orient during migration.
Recent advancements are the identification of a suitable protein, which is present in
the retina and possesses the properties required by the avian magneto-reception \cite{Ritz00,Ritz09},
and the experimental demonstration that a RP reaction can be sensitive to 
magnetic fields as weak as the Earth's one \cite{Maeda08}.
The formulation of an additional quantum physics perspective boosted
the interest towards the study of entanglement, its decoherence and the implementation
of quantum control techniques \cite{GG10,Vedral11,Cai11a,Cai11b}.
In this respect, a RP reaction constitutes an electron spin system exposed to
a mesoscopic, non-Markovian nuclear spin bath.

The theory describing the radical pair chemical reactions
has been developed in the spin chemistry community and has successfully undergone
a series of experimental tests \cite{Buchachenko,Turro}.
Still, it would be necessary to investigate the model on the short timescale
of the electron spin dynamics.
Here, we propose a new experimental setup which allows sub-nanosecond
time resolution by effectively engineering the reaction kinetics.
We suggest to connect the radicals via a molecular switch presenting two isomeric
forms and capable of undergoing conformational changes upon the absorption of
a photon.
In this way, one can optically manipulate the distance between the radicals
and effectively determine the instant of their re-encounter.
Since the chemical reaction takes place at the moment of re-encounter,
we achieve control on the reaction kinetics.
This new possibility adds in a complementary way to previous proposals
in which the electron spin dynamics is influenced by the application
of magnetic pulses \cite{GG10} or oscillating fields \cite{Cai11b}.

First, we present the experimental setup and discuss its feasibility with state-of-the-art
technology. The consistency of time scales and the efficiency of the procedure are
analyzed introducing the case study of azobenzene as the photo-controlled bridge.
Second, we show how the additional control can be exploited to improve the performances
of sensors based on the RP mechanism, and in particular we apply our scheme to a chemical magnetometer
to boost its sensitivity by two orders of magnitude. It is remarkable that such enhancement
is even obtained using a conservative, \textit{i.e.} coarse-grained, time resolved scheme.

\textit{Radical Pair Chemical reactions.---}
The name RP reactions refers to a broad class of chemical reactions
characterized by the creation of radicals
as intermediate products \cite{Steiner89}.
Here, we consider those reactions which take place in solution and
involve two different chemical compounds, let us call them $A$ and $D$ for acceptor and
donor respectively, free to diffuse.
Typically, one of the molecules is photoexcited and, if $A$ and $D$ are close enough,
a fast electron transfer establishes the radical pair
D$^{\bullet +}$-A$^{\bullet -}$.
The electron spins are assumed to be in the singlet state
$\ket{S}=\frac{1}{\sqrt{2}}(\ket{\uparrow\downarrow}-\ket{\downarrow\uparrow})$,
while, at room temperature, the nuclear spins are in the completely mixed state.
The radicals diffuse away from each other to such distances that the electron-electron
interactions become negligible and the electron spin dynamics is determined only by
the external magnetic field and the interaction of each electron with
the surrounding nuclear spins on the respective radical \cite{Steiner89}:
\begin{equation}
H=\sum_{k=1,2}H_k=-\gamma _e\vec{B}\cdot
\sum_{k}\vec{S}_k + | \gamma _e | \sum_{k,j} \vec{S}_k\cdot \hat{\alpha}_{k_j}
\cdot \vec{I}_{k_j}  \label{Hamiltonian}
\end{equation}
where $\gamma _{e}=-g_e\mu_B$ is the electron gyromagnetic
ratio, $\vec{S}_k$, $\vec{I}_{k_j}$ are the electron and nuclear
spin operators respectively, $\vec{B}$ is the external magnetic field,
$\hat{\alpha}_{k_j}$ denote the hyperfine coupling tensors,
and we fix $\hbar=1$.
In a stochastic way, geminate re-encounter of the two radicals happens, and
backward electron transfer completes the chemical reaction for those radicals
in a singlet state.
Triplet radicals recombine to distinct triplet products.

The number of radical pairs which recombine through the singlet
reaction channel is directly proportional to the time integral of the
fraction of radicals in a singlet state $f_s(t)$ weighted by the probability of
(diffusive) re-encounter:
$ \Phi_{s}=\int_{0}^{\infty}p_{re}(t) f_s(t) \text{d}t $ ,
where $f_s(t)=\bra{S} \rho_{el}(t) \ket{S}$ is the overlap between the singlet state
and the electron spin state $\rho_{el}(t)$ at time $t$.
Thus, we are facing a system in which two distinct, but concurrent dynamics are
taking place: The quantum evolution of the electron spin state, and the classical
diffusion of the molecules D$^{\bullet +}$ and A$^{\bullet -}$. The latter one
stochastically determines the duration of the former.
In the literature, the probability distribution for the re-encounter time
has phenomenologically been described as a single exponential
$p_{re}(t)= k e^{-k t}$ \cite{Steiner89,Rodgers07}
with re-encounter rate $k$.
It is $p_{re}(t)$ that we aim to control in order to modify $\Phi_{s}$
and all the related quantities.

\begin{figure}[b]
 \centering
 \subfigure[flexible chain]
   {\includegraphics[width=2.3cm]{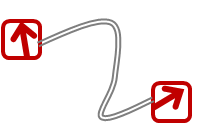}}
 \hspace{15mm}
 \subfigure[micelle]
   {\includegraphics[width=1.8cm]{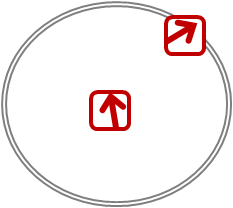}}
 \hspace{15mm}
 \subfigure[photoswitching bridge]
   {\includegraphics[width=4.7cm]{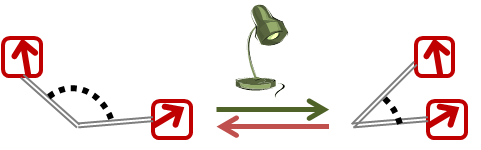}}
 \caption{Three approaches to modify the re-encounter probability:
 (a) The radicals are linked by a flexible, inextensible chain,
 (b) the chemical reaction happens inside a micelle, (c) the radicals
 are linked by a rigid, optically controlled bridge.}
 \label{approaches}
\end{figure}

\textit{Re-encounter Probability.---}
A large rate $k\sim 1\,$ns$^{-1}$ indicates that most
of the geminate re-encounters happen in the first few nanoseconds after
the initial electron transfer, and so it becomes hard to study the
electron spin dynamics at later times or to observe the dependence
of the entanglement lifetime on $B$ as predicted in \cite{GG10}.
Furthermore, a weak magnetic field effectively has no time to produce
any appreciable effect and this limits the sensitivity of chemical
magnetometers or compasses.
In the context of avian magneto-reception, the RP is actually required
\cite{Ritz00,Rodgers09} to have a small re-encounter rate $k \sim 1\,\mu$s$^{-1}$.
Unfortunately this is generally not the case for chemical reactions
involving bi-ionic radicals in solution since the Coulomb attraction
leads to a fast re-encounter.
Thus several experimental efforts have been undertaken to modify the
distribution of the RP re-encounter time, mainly by imposing constraints
on the geometry of the system as shown in Fig.~\ref{approaches}:
(a) The radicals are linked by a flexible, inextensible chain
whose length determines the maximum separation that they can achieve \cite{Enjo97};
(b) The reaction takes place inside a micelle which effectively defines
a much smaller container where the radicals are allowed to diffuse \cite{Eveson00}.
The experimental studies of RP reactions in these systems confirm
a substantial modification of the re-encounter time distribution, but are still
interpreted within the phenomenological exponential model
with rate $k\sim 0.1-1 \,\mu$s$^{-1}$ essentially determined by solvent viscosity,
chain length, hydrophobic tail of the surfactant, and size of the micelles.
These properties cannot in general be changed continuously or even independently,
and their effect on the rate $k$ is not easy to predict quantitatively.
To overcome these difficulties, we propose to use optical switches, in this way
combining geometrical constraints with the possibility of optically controlling them.

\textit{Photo-controlled radicals.---}
The distance of the radicals strongly affects both the possibility of recombination
and the electron spin dynamics.
At small separations, the electron transfer is favoured while the electron
spin dynamics is suppressed due to the large exchange and dipolar interaction;
conversely, at large distances, the electron transfer is inhibited
while the dynamics is determined by the Zeeman and hyperfine interaction only,
since the exchange and dipolar interactions are now negligibly weak.
We propose to optically control the inter-radical separation by chemically
attaching $A$ and $D$ to the two endings of a molecule exhibiting
two isomeric forms and an isomerization pathway which is photo-activated.
Such molecules possess the property of reversibly switching between a straight
and a contracted structure upon the absorption of a photon
of suitable wavelength.
The two isomeric forms effectively impose a ``closed'', respectively ``open'',
configuration for the radicals, see Fig.~\ref{approaches}, that
we are able to change in a reversible way by means of short laser pulses.
Examples of photoswitchable compounds are fulgides,
diarylethenes or azobenzene derivatives \cite{Tamai00}.

For this description to hold, it is necessary that the photo-isomerization process
is effective and causes a substantial change of the distance between the two radicals.
In this case, and via the optical manipulation of the opening and closing
of the linking bridge, it is in principle possible to directly control
the instant of RP creation, the exact duration of the quantum dynamics for
the electron spins, and the instant of the final recombination.
The implementation of such a scheme is ideally as follows:
Initially all the bridges are closed and the radicals not yet formed,
then the RP are created by excitation via a laser pulse (standard technique
applied for example in \cite{Maeda08}), and immediately after that, a second
laser pulse triggers the isomerization and separates the radicals.
The electron spin state evolves according to the Hamiltonian in
Eq.~(\ref{Hamiltonian}) until a third laser pulse closes the bridges and
leads to the final recombination.
It is desirable, but not strictly necessary, that all the three processes could be
selectively triggered by laser pulses of different wavelengths.

Under the condition that the duration of the various pulses remains much shorter
than the time scale of the electron spin evolution,
the re-encounter probability can be described by a peaked distribution
$p_{re}(t) \propto \delta (t-\tau_{re})$, with $\tau_{re}$ being the time elapsed
between the second and third pulse and ``$\propto$'' indicating
the proportionality with respect to the actual photo-isomerization yield.
Such peaked distribution allows to monitor the electron spin dynamics at
the precise instant $\tau_{re}$ which can be chosen without restrictions
in a wide interval ranging from a few nanoseconds to several tens of microseconds
(for later times one has to take into account spin-lattice relaxation).
The proposed scheme can be straightforwardly adapted to engineer more
complicated and structured re-encounter probability distributions:
A series of pulses creates a re-encounter probability exhibiting separated peaks,
while continuous illumination gives rise to an exponential
distribution (here not phenomenologically assumed, but externally imposed)
whose rate is directly proportional to the light intensity.
Combining laser pulses, dark intervals, and continuous illumination with
adjustable light intensity, the ability of engineering a
specific $p_{re}(t)$ is practically limited by technical restrictions
(\textit{e.g.} the laser intensity or the speed of light modulation)
and by the absorption and isomerization properties of the photoswitches
(\textit{e.g.} the selectivity with which an ``opening pulse'' does not cause
spurious ``closing events'', and viceversa).
Realistically, a fraction of the photoswitches may not undergo
photoisomerization, but, for thermally stable isomers, this means
that the corresponding radicals simply do not contribute to the
chemical reaction.
The actual isomerization yield determines the number of radical
pairs effectively involved in the experiment.

Note, that an alternative scenario is given by photoswitches that do
not change their conformation, but rather change their conductance, passing
from an insulating to a conductive state and viceversa \cite{DelValle07,vanderMolen10}.
In the case of backward electron transfer through the linker, the isomerization
of the linker corresponds to switching on/off the coupling between the two radicals
at its ends.

\textit{Case study: azobenzene.---}
To explore the actual feasibility of our proposal, we consider the
concrete example of azobenzene as the photoswitching molecular bridge.
Azobenzene is an organic molecule composed of two phenyl rings linked
by a N$=$N double bond, and occurs in two isomeric forms, trans and cis
azobenzene, having a stretched, respectively contracted, structure.
The synthesis of azobenzene derivatives having different chemical functional
groups extending from the phenyl rings is routinely performed
\cite{Hugel02,Hofmann09}.
Preliminary checks have to verify that the chemistry of the
radical pair is not drastically altered by the presence of the
azobenzene and that the constraints on the RP distance are satisfied.

For a realistic implementation of our proposal, the involved
processes need to satisfy a hierarchy of time scales:
The electron spin dynamics determines the reference time scale
to be $\sim 1-10\,$ns (for typical values of the external magnetic field
and hyperfine couplings, \textit{i.e.} $\sim 0.1-1\,$mT) with respect
to which the spin lattice relaxation and the thermally driven isomerization
have to be slow, while the radical creation and photoisomerization
mechanism have to be fast.
Typically, the spin-lattice relaxation takes $\sim 1-10\,\mu$s
at room temperature, while the donor photo-excitation and successive
electron transfer happens in a few picoseconds.
In the case of azobenzene, the isomers can be considered completely
stable at room temperature on the time scale of seconds \cite{Rau82},
while experimental studies with azobenzene in solution afforded S$_1$ excited state life times
of trans-azobenzene of $\sim 2.5\,$ps when excited without excess vibrational
energy \cite{Lednev98} and $\sim 0.5\,$ps when from higher vibrational states
of S$_1$ populated after internal conversion from S$_2$ \cite{Fujino01}.
Finally, the time scale over which we can modulate the re-encounter
probability depends on the duration of the optical pulses:
The estimates in \cite{Yamaguchi09} suggest that a single pulse of $150\,$fs FWHM
at $439\,$nm 
and having an energy of $200\,$nJ leads to $3\%$
cis-to-trans photoisomerization yield.
A single pulse of duration $\simeq 10$ ps and analogous intensity can achieve
photoisomerization yield of $30-40\%$.
Concerning the process of RP creation, in reference \cite{Maeda08}
radicals were generated by a $7\,$ns , $5\,$mJ pulse, but even radical
pair generation in less than one picosecond has been achieved
in experiments with femtosecond laser pulses \cite{Gilch98}.

\textit{Application: chemical magnetometry.---}
The control of the RP re-encounter probability finds immediate
application to increase the performance of chemical devices.
Here, we show how a simple-to-implement control scheme
highly enhances the sensitivity of a chemical magnetometer
by two orders of magnitude.
The basic idea behind the functioning of a chemical magnetometer
is that, since a change in the magnetic field modifies the amount
of singlet products, one can reverse the reasoning and measure the
chemical yield to estimate $B$.
By definition, the magnetic sensitivity is high when a small
change in the magnetic field intensity produces large effects
on the singlet yield:
\begin{equation}
\Lambda_s(B) \equiv \frac{\partial\Phi_{s}(B)}{\partial B} =
                \int_{0}^{\infty}p_{re}(t) g_s(B,t) \text{d}t,
\label{sensitivity}
\end{equation}
with $g_s(B,t) \equiv \frac{\partial f_s(B,t)}{\partial B}$
being the instantaneous magnetic sensitivity.
The functional form of $f_s(B,t)=\bra{S} \rho_{el}(t) \ket{S}$ strongly
depends on the specific realization of the radical pair, in particular
on the number of the surrounding nuclear spins.
Here, we consider a radical pair in which the first electron spin is
void of hyperfine interactions, while the second electron spin
interacts isotropically with one spin-1 nucleus.
In the context of the chemical compass (i.e. when
the task is determining the magnetic field direction through anisotropic
hyperfine interactions), an analogous configuration (with only one spin-1/2
nucleus) has been proposed \cite{Ritz09}, and numerically characterized
\cite{Cai11b}, as being optimal:
Additional nuclear spins would perturb the intuitive
``reference and probe'' picture.
The Hamiltonian then simplifies to
$H = -\gamma _e B (S_1^{(z)}+S_2^{(z)}) + | \gamma _e | \alpha \vec{S}_2 \cdot \vec{I}$,
where $\alpha$ is the isotropic hyperfine coupling.

\begin{figure}[t]
 \centering
   \includegraphics[width=8cm]{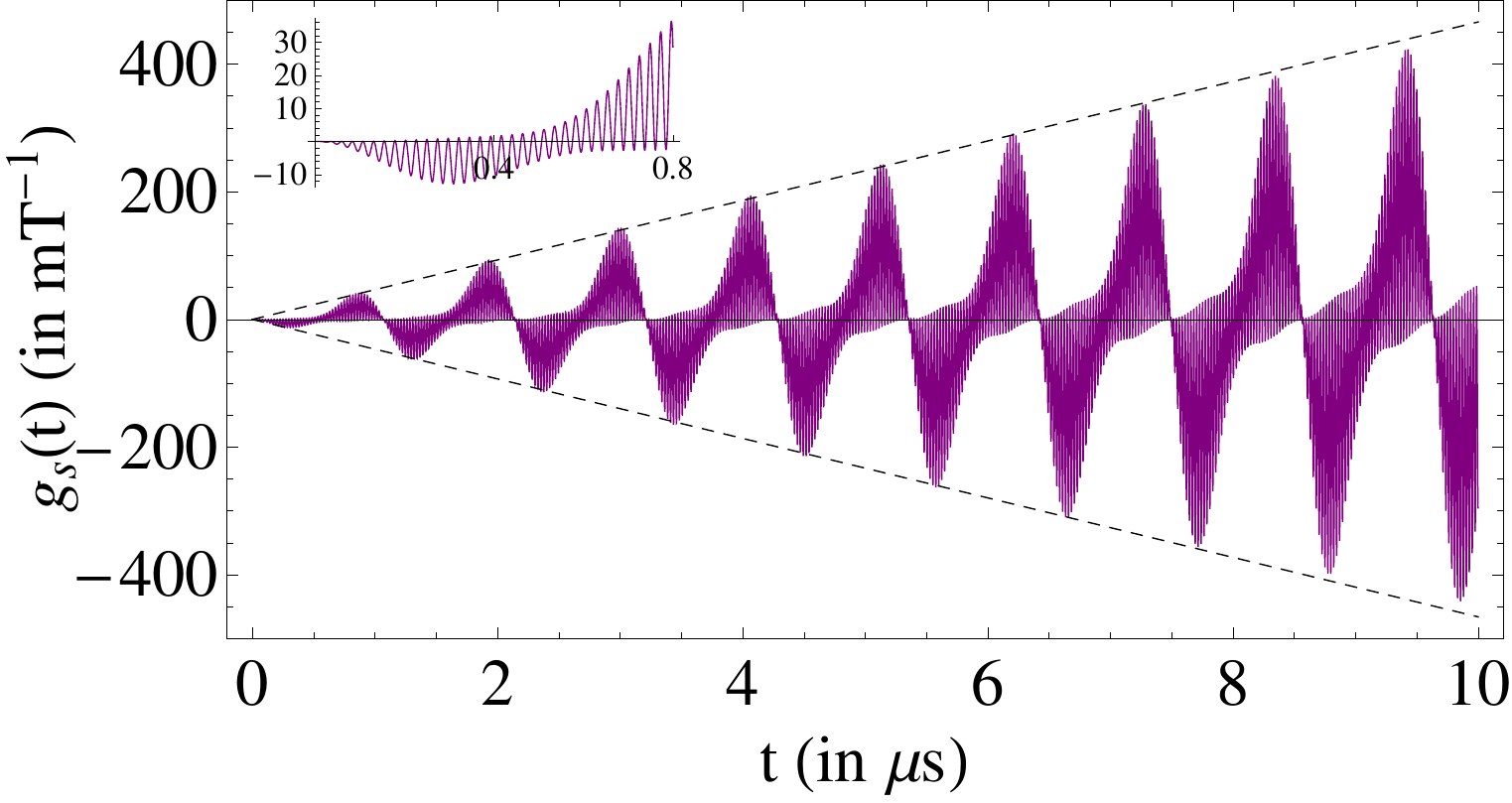}
 \vspace{-2mm}
 \caption{Instantaneous magnetic sensitivity for $\alpha=1$ mT and
 $B=\frac{\alpha}{20}$. The peaks of the envelope grow linearly with time and
 the envelope's shape is approximately periodic with period
 $\tau_{envel}=\frac{3 \pi}{|\gamma_e|B}\approx 1.07\,\mu$s.
 Inside the envelope, $g_s(t)$ oscillates with period
$\simeq \frac{4 \pi}{3|\gamma_e|\alpha}\approx 23.8\,$ns (see inset).}
\label{g_s}
\end{figure}

We evaluate the sensitivity in the regime of long spin-lattice relaxation
times and weak magnetic fields, \textit{i.e.} for $B\ll \alpha$, noting
that in practical applications one can reproduce such condition
by applying a suitable external field for compensation.
For the realistic choice $\alpha \simeq 1.0\,$mT, the Earth magnetic
field $B_{Earth} \simeq \alpha/20$ is indeed weak.
Straightforward, but rather tedious, calculations lead to an analytical
expression for the instantaneous magnetic sensitivity $g_s(B,t)$ \cite{supplementary},
whose behavior, in the regime $B\ll \alpha$, is shown in Fig.~\ref{g_s}.

\begin{figure}[]
 \centering
   \includegraphics[width=4.1cm]{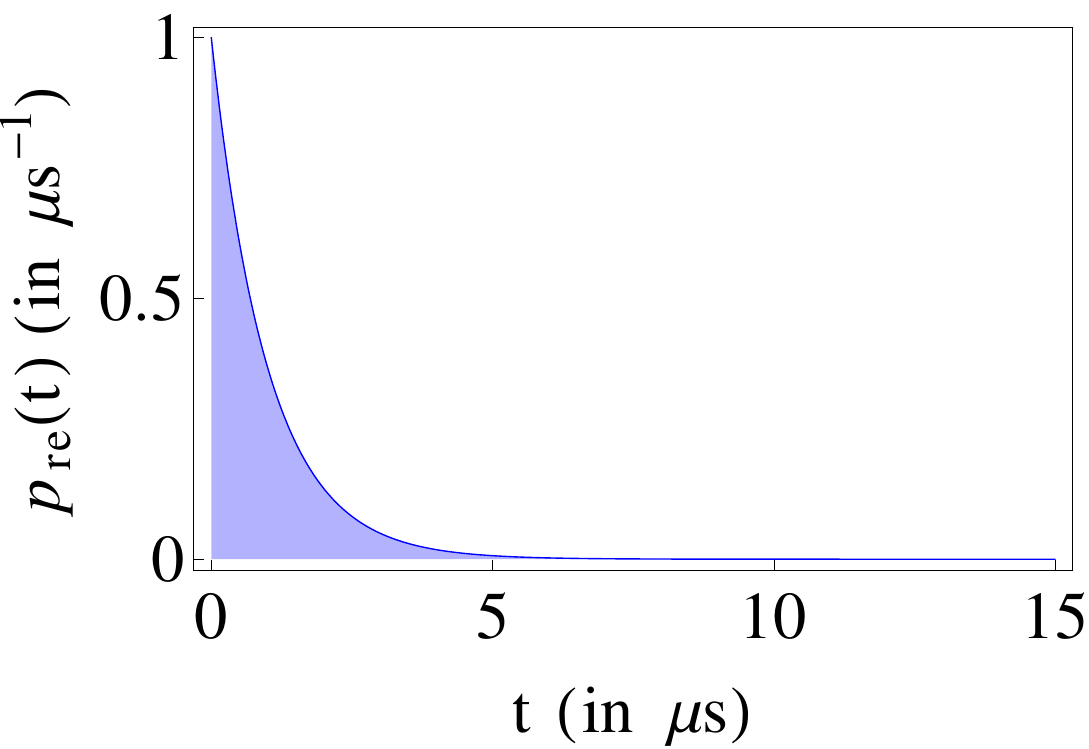}
 \hspace{2mm}
   \includegraphics[width=4.1cm]{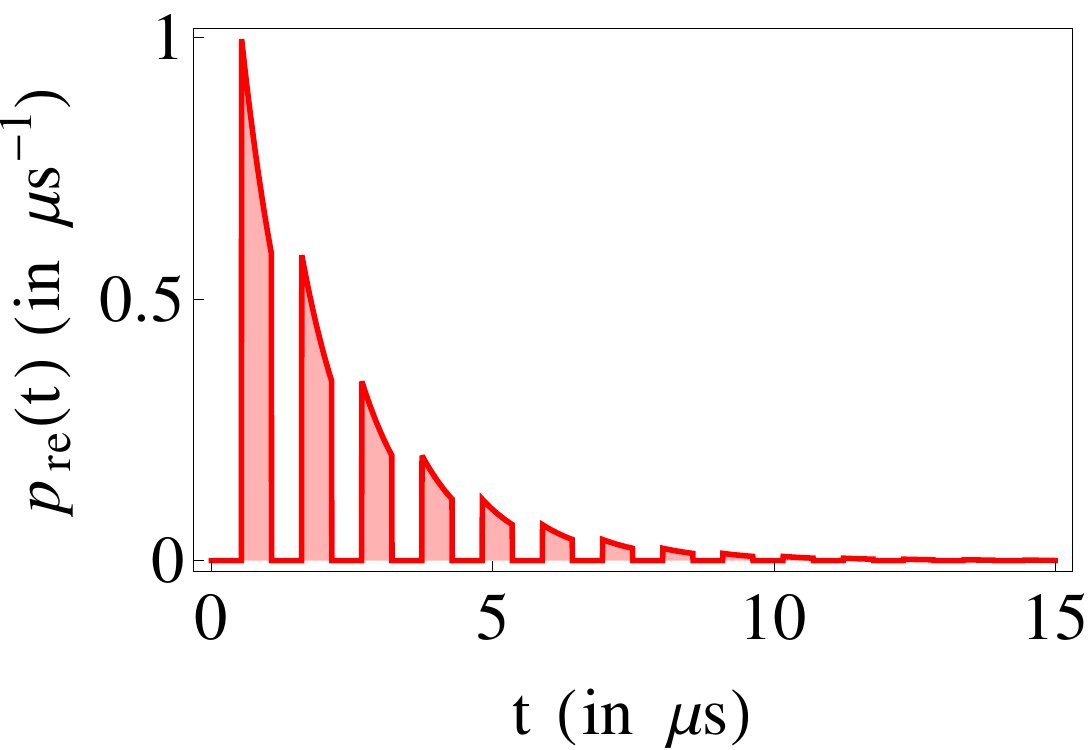}
 \\
   \includegraphics[width=4.1cm]{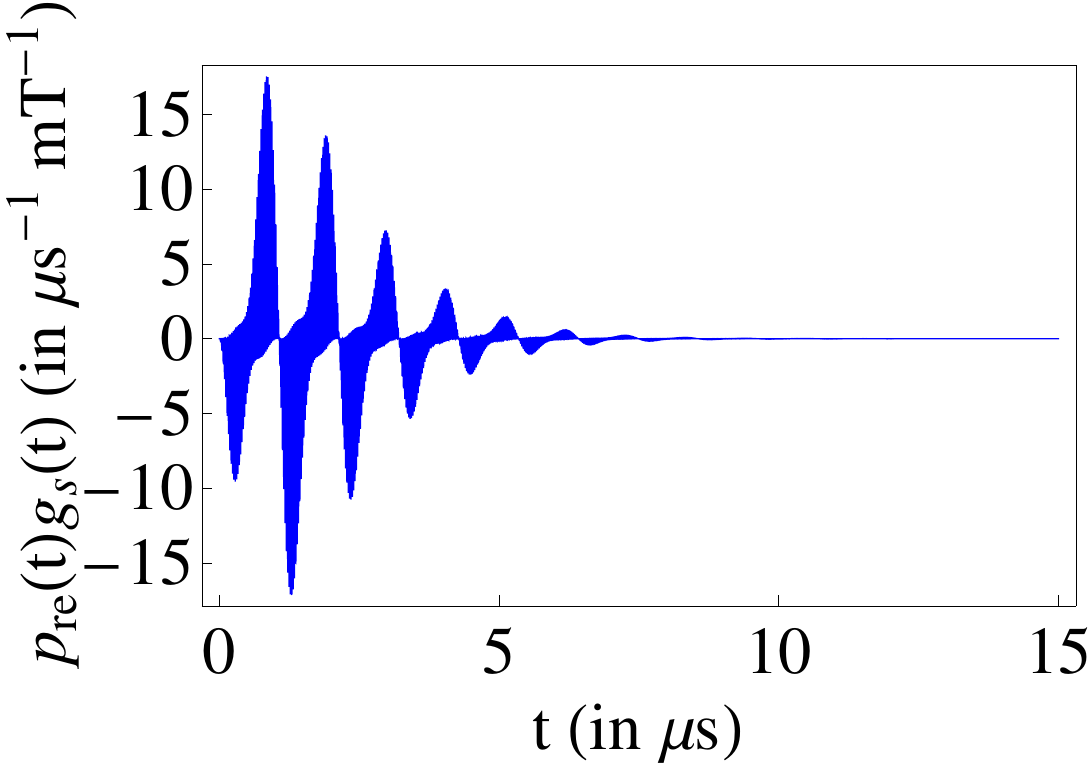}
 \hspace{2mm}
   \includegraphics[width=4.1cm]{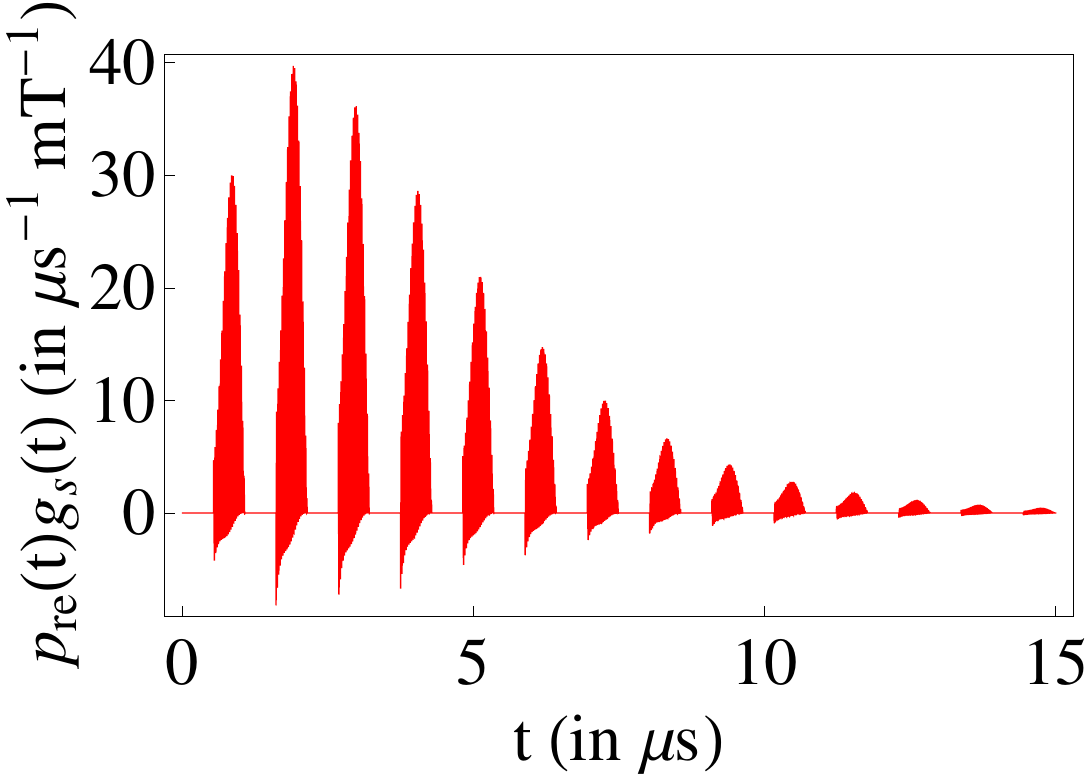}
\caption{Re-encounter probability (top) and integrand for
the magnetic sensitivity (bottom) over time. The blue curve (left)
corresponds to the phenomenological exponential model,
the red curve (right) to the optically controlled case.
Note that we have chosen the laser intensity such that
$k_{prot}=k=1 \mu$s$^{-1}$ to make the comparison more direct.
In general, while $k$ is determined by the radicals and properties
of the solution, $k_{prot}$ is a tunable parameter of the protocol.
}
\label{g*p}
\end{figure}

When integrated on timescales longer than the effective ``period'' of the
envelope, $\tau_{envel}=\frac{3\pi}{|\gamma_e|B}$,
the positive and negative contributions cancel each other and
therefore drastically reduce the ultimate sensitivity $\Lambda_s(B)$.
A possible solution is to engineer the re-encounter probability
in such a way that the radical pairs are allowed to recombine only
when $g_s(B,t)\geq0$.
Such desirable $p_{re}(t)$ is obtained applying weak light pulses
(for the final step of ``closing'' the photoswitchable bridge)
interspersed with equally long dark intervals, both of duration
$\tau \simeq \tau_{envel}/2$.
Considering the realistic parameters of Fig.~\ref{g_s}, one
obtains $\tau \approx 0.54\,\mu$s corresponding to a laser repetition
of around $1\,$MHz.
The corresponding re-encounter probability is a piecewise exponentially
decreasing function with decay rate $k_{prot}$ directly proportional
to the light intensity.
The phenomenological and controlled re-encounter probabilities are
shown in Fig.~\ref{g*p} together with the corresponding integrand
involved in the evaluation of the magnetic sensitivity via Eq.~(\ref{sensitivity}).

\begin{figure}[b]
 \centering
   \includegraphics[width=4.2cm]{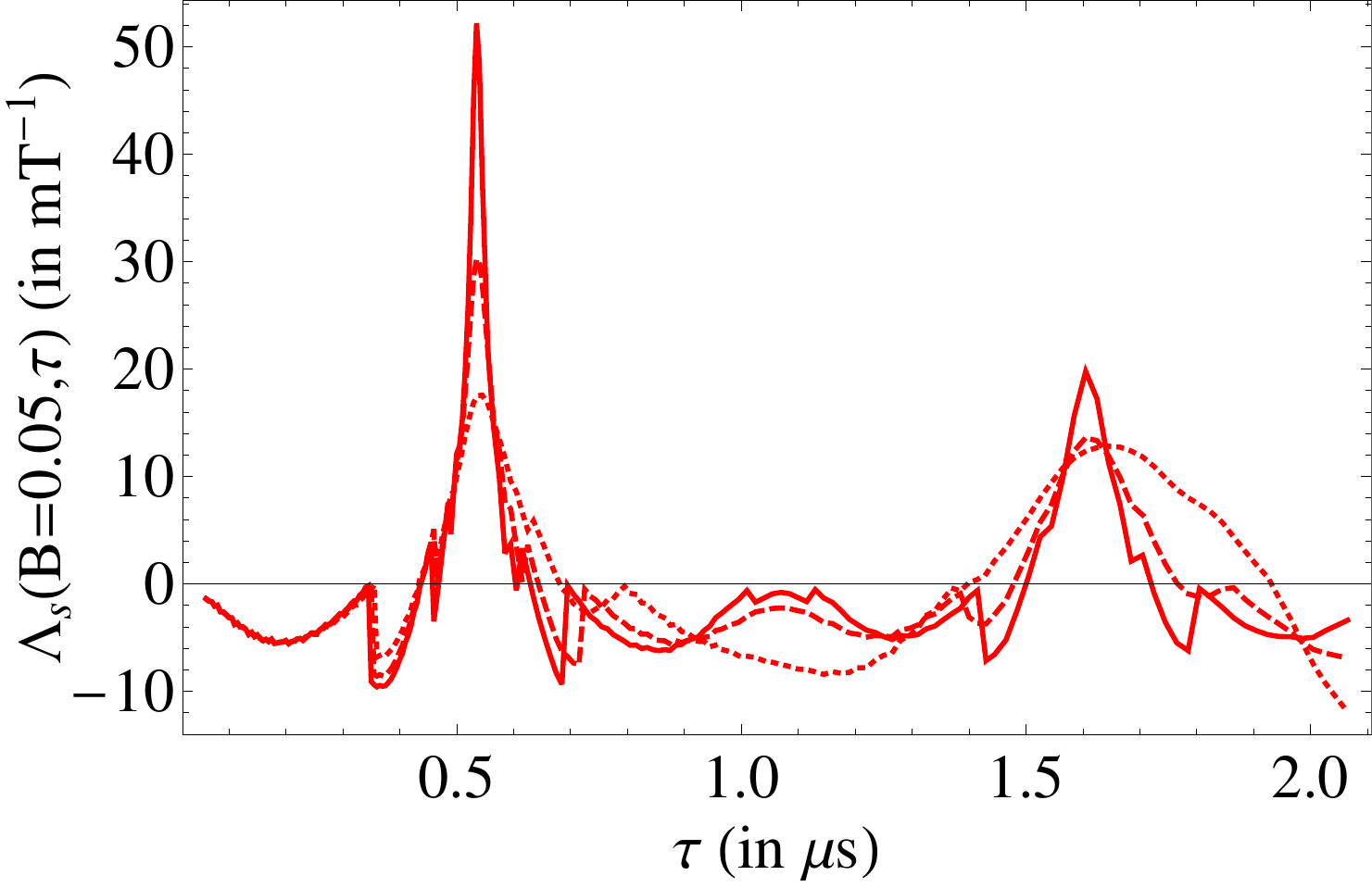}
 \hspace{0mm}
   \includegraphics[width=4.2cm]{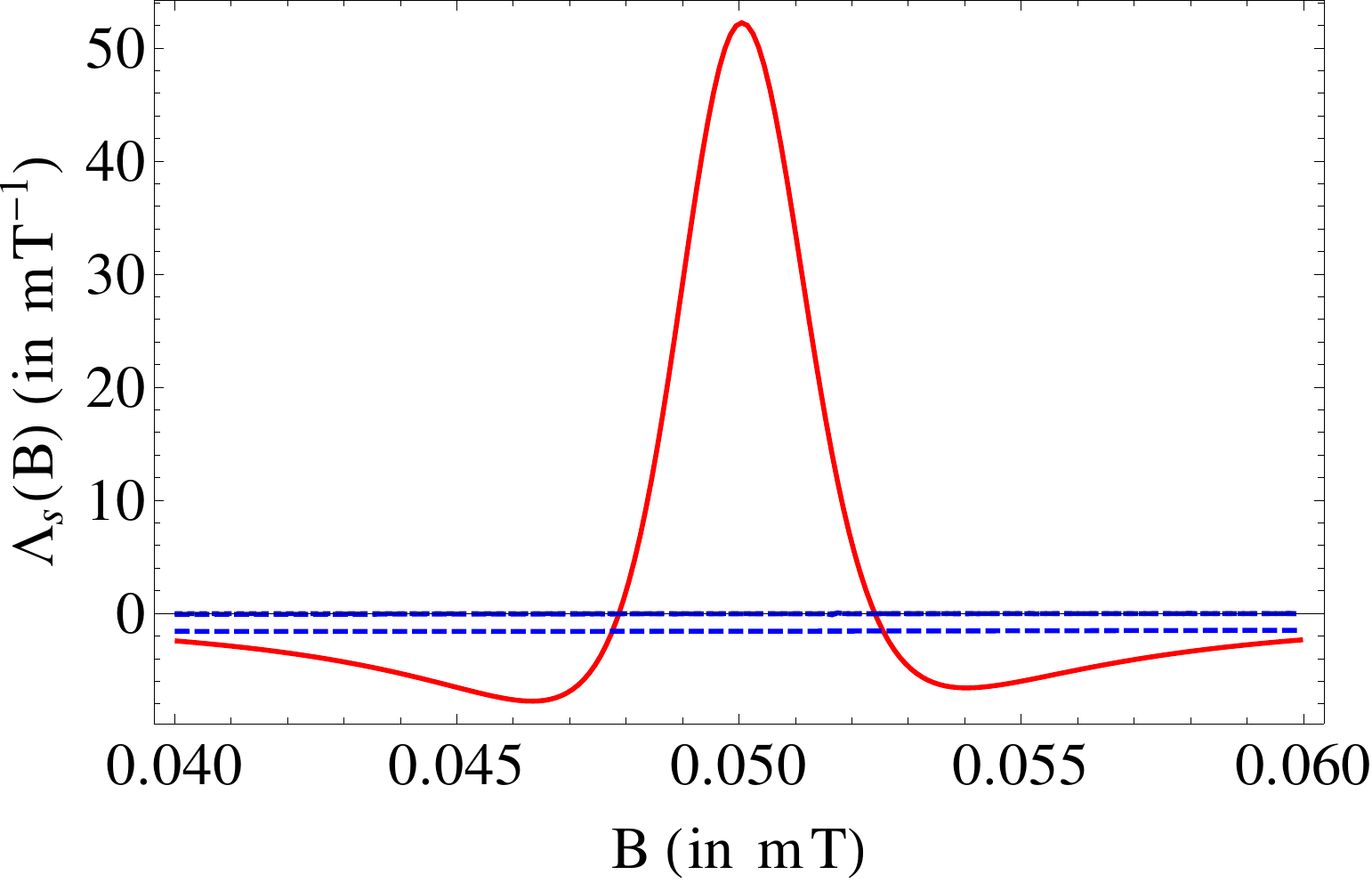}
\caption{Left: Magnetic sensitivity $\Lambda_s$ \textit{vs.} pulse duration $\tau$.
 $B=0.05\,$mT and $1/k_{prot}=2\,\mu$s (solid), $1\,\mu$s (dashed), $0.5\,\mu$s (dotted).
 Right: Magnetic sensitivity $\Lambda_s$ \textit{vs.} field intensity $B$
 for the exponential model with $1/k=10\,$ns (solid, blue), $1/k=100\,$ns (dashed, blue),
 $1/k=2\,\mu$s (dotted, blue), and for our proposal with pulse duration $\tau^\prime=0.535\,\mu$s
 (solid, red). Parameters as in Fig.~\ref{g_s}.
}
\label{results}
\end{figure}

The change in the (integrated) magnetic sensitivity is very sensitive
to the duration of the pulses, so we propose to monitor
$\Lambda_s(B,\tau)$ (as currently done in spin chemistry experiments
\cite{Kalneus05,Grampp05}) while changing the pulse duration $\tau$.
From an irregular landscape, a sharp peak centered around
a certain $\tau^\prime$ emerges due to the resonant condition
$\tau^\prime\simeq \tau_{envel}/2$, see Fig.~\ref{results} (left)
in which $\tau^\prime=0.535\,\mu$s.
As expected, at $\tau\simeq 3\tau^\prime$,
a weaker resonance gives rise to a second peak.
Thus, a first estimate of the field intensity is given by
$B_{estim}=\frac{3 \pi}{2 |\gamma_e|\tau^\prime}$, and its precise
value is obtained by applying pulses of duration $\tau^\prime$.
Fig.~\ref{results} (right) shows that the optical control
of the RP re-encounter probability enhances the magnetic
sensitivity up to two orders of magnitude
with respect to the phenomenological exponential model
over a broad range of rate $k$.
A similar enhancement is obtained for several $k_{prot}$
satisfying $k_{prot} \leq \tau^{-1}$.
Similar results are obtained for a single nuclear spin-1/2
and a suitably modified protocol \cite{supplementary}.

\textit{Summary and Outlook.---}
We have proposed a new experimental approach which enables
the control of the re-encounter probability of radical
pairs in solution via light pulses.
This will allow one to study the electron spin dynamics on the short
time scale of its evolution and thus enlarge the information
accessible in spin chemistry experiments.
The control of the reaction kinetics constitutes a complementary
and autonomous method compared to previous works proposing
direct electron spin manipulation \cite{GG10,Cai11a,Cai11b}.
Such an additional handle on the RP dynamics should contribute to a
better understanding of the radical pair mechanism itself.

In order to improve the performance of man-made sensors
based on chemical reactions, one naturally pursues the approach
of artificially synthesizing suitable compounds like, in our proposal,
radical pair molecules connected to a photoswitchable bridge.
As a concrete result, we have presented a simple protocol
which enhances the sensitivity of a chemical magnetometer
by up to two orders of magnitude.
Remarkably, this scheme can be implemented using weak laser
pulses having a duration much longer than the time scale given 
by the electron spin dynamics.

\textit{Acknowledgements.---}
This research was supported by the Austrian Science Fund (FWF): F04011, F04012.

{}

\end{document}